\documentclass[journal]{IEEEtran}

\ifCLASSINFOpdf
\else
   \usepackage[dvips]{graphicx}
\fi
\usepackage{url}

\usepackage{bm}
\usepackage{cite}
\usepackage{xcolor}
\usepackage{pifont}
\usepackage{amsthm}
\usepackage{comment}
\usepackage{amsmath}
\usepackage{amssymb}
\usepackage{balance}
\usepackage{graphicx}
\usepackage{amsfonts}
\usepackage{arydshln}
\usepackage{graphicx}
\usepackage{setspace}
\usepackage{algorithm}
\usepackage{algorithmic}

\newtheorem{lemma}{Lemma}
\newtheorem*{lemma*}{Lemma}

\newtheorem{theorem}{Theorem}
\newtheorem*{theorem*}{Theorem}

\newtheorem{definition}{Definition}
\newtheorem{proposition}{Proposition}

\def\BibTeX{{\rm B\kern-.05em{\sc i\kern-.025em b}\kern-.08em
		T\kern-.1667em\lower.7ex\hbox{E}\kern-.125emX}}

\begin{document}

\title{Asymptotic Achievability of the Cramér-Rao Lower Bound of Channel Estimation for Reconfigurable Intelligent Surface Aided Communication Systems
\thanks{This work is supported by the National Natural Science Foundation of \textsc{China} (No. 4217010675, 61771345), Science and Technology Commission of \textsc{Shanghai} Municipality (No. 19511102002), and \textsc{Shanghai} Technical Innovation Action Plan (No. 21220713100).  \emph{Corresponding Author: Erwu Liu}.
}
}

\author{Yiming~Liu,~\IEEEmembership{Graduate Student~Member,~IEEE,}
		Erwu~Liu,~\IEEEmembership{Senior~Member,~IEEE,}\\
		Rui~Wang,~\IEEEmembership{Senior~Member,~IEEE,} 
		Zhu~Han,~\IEEEmembership{Fellow,~IEEE,}
		Binyu~Lu,~\IEEEmembership{Graduate Student~Member,~IEEE}
	
		\thanks{
			Yiming Liu, Erwu Liu, Rui Wang, and Binyu Lu are with the Department of Information and Communication Engineering, Tongji University, Shanghai 201804, China, E-mail: ymliu\_970131@tongji.edu.cn, erwu.liu@ieee.org.}
		\thanks{
			Zhu Han is with the Department of Electrical and Computer Engineering, University of Houston, Houston, TX 77004 USA, E-mail: Zhan2@uh.edu.}
}

\maketitle

\begin{abstract}
	 To achieve the joint active and passive beamforming gains in the reconfigurable intelligent surface assisted millimeter wave system, the reflected cascade channel needs to be accurately estimated. Many strategies have been proposed in the literature to solve this issue. However, whether the Cramér-Rao lower bound (CRLB) of such estimation is achievable still remains uncertain. To fill this gap, we first convert the channel estimation problem into a sparse signal recovery problem by utilizing the properties of discrete Fourier transform matrix and Kronecker product. Then, a joint typicality-based estimator is utilized to carry out the signal recovery task. We show that, through both mathematical proofs and numerical simulations, the solution proposed in this letter can asymptotically achieve the CRLB.	
\end{abstract}


\IEEEpeerreviewmaketitle

\section{Introduction}

\IEEEPARstart{R}{econfigurable} intelligent surface (RIS) technology is a very promising and cost-effective solution to improve the spectrum and energy efficiency of wireless communication systems {\cite{8936989, 8741198, Liu2012:Energy, 9366805}}. With the assistance of a smart controller, the RIS can adjust its reflection coefficients such that the desired signals are added constructively. The joint active and passive beamforming design has been studied in many existing works with continuous phase shifts (e.g., {\cite{8811733, 9133184}}) or discrete phase shifts (e.g., {\cite{8930608, 9013288}}) at reflecting elements. Moreover, the RIS also can be used in millimeter wave (mmWave) systems {\cite{8485924}}. 

To achieve the above joint active and passive beamforming gains, the cascade channel should be estimated efficiently and accurately. However, in the scenario of mmWave channels, it is difficult to establish a scheme that can simultaneously achieve high accuracy and efficiency. In consequence, efficiency is the priority in the existing work. Several novel strategies have been proposed to efficiently make such estimations. Authors in {\cite{8611231}} utilized the generalized approximate message passing (GAMP) algorithm to find the entries of the unknown mmWave channel matrix. Similarly, in {\cite{9133156}}, authors adopted the message passing (MP) based algorithm to estimate the cascade channels. The orthogonal matching pursuit (OMP) method was used in {\cite{9103231}}. Nevertheless, the existing schemes cannot achieve the optimal estimation accuracy, i.e., the Cramér-Rao lower bound (CRLB) of channel estimation for RIS-aided mmWave systems.

Contrary to these efficient algorithms, we intend to establish a scheme which can achieve the CRLB. For this purpose, we first convert the channel estimation task into a noisy sparse signal recovery problem through utilizing the properties of the discrete Fourier transform (DFT) matrix and the Kronecker product. 
Then, a joint typicality-based estimator is proposed to carry out the recovery task and establish the asymptotic achievability of the CRLB when the product of the number of receiver antennas and the number of time slots approaches infinity. 
The correctness of our result is verified through both mathematical proofs and numerical simulations.
In addition, based on the sparsity structure established in this letter, our analysis result can also be applied to the conventional point-to-point mmWave system which is a special case of RIS-assisted systems. However, it should be noted that although it is the first result establishing the achievability of the CRLB of channel estimation for RIS-assisted mmWave systems, our scheme is complex and costs a lot of overhead. Thus, finding a lower-complexity estimator that can simultaneously achieve the CRLB is the future important work.

\section{System and Channel Model}

\subsection{System Model}
 We consider an RIS-assisted mmWave  system, as illustrated in \textcolor{black}{Fig. 1}, where the base station (BS) and the mobile station (MS) are equipped with $N_{\mathrm{s}}$ and $N_{\mathrm{d}}$ antennas, respectively, and the RIS is equipped with $N_{\mathrm{r}}$ reflecting elements.
Although the BS and the MS are equipped with a large number of antennas, they can fit within the compact form because of the small wavelength of mmWave. 
In this letter, to better illustrate our results, we neglect the direct link from the BS to the MS. Nevertheless, the extension to the scenario with the direct link is straightforward. 
In addition, due to the inherent sparsity of mmWave channels {\cite{6834753}}, there exists only a dominant line-of-sight path and very few non-line-of-sight paths in the BS-RIS link and the RIS-MS link.
Then, the elevation (azimuth) angle-of-departure (AoD) of the $i^{\textit{th}}$ path at the BS and the RIS are denoted as $\theta_{i}$ ($\phi_{i}$) and $\gamma_{i}^{\prime}$ ($\mu_{i}^{\prime}$), respectively, and the elevation (azimuth) angle-of-arrival (AoA) of the $i^{\textit{th}}$ path at the RIS and the MS are denoted as $\gamma_{i}$ ($\mu_{i}$) and $\vartheta_{i}$ ($\varphi_{i}$), respectively. 
\begin{figure}[htbp]
	\centerline{\includegraphics[width = 7 cm]{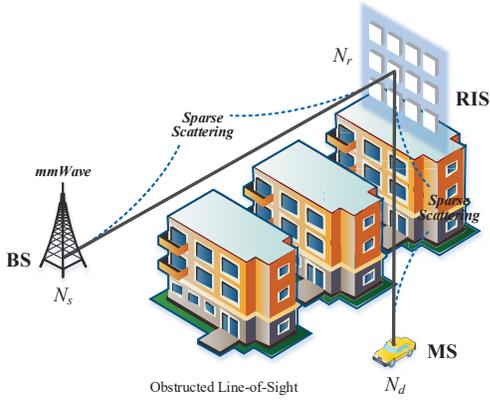}}
	\caption{The RIS-assisted mmWave communication system with an $N_{\mathrm{s}}$-antenna BS, an $N_{\mathrm{d}}$-antenna MS, and an RIS comprising $N_{\mathrm{r}}$ reflecting elements.}
	\label{fig 1}
	\vspace{-13 pt}
\end{figure}

\subsection{Channel Model}

Due to the inherent sparse nature of mmWave channels, the number of paths between the BS and RIS is small relative to the dimensions of BS-RIS channel matrix ${\mathbf{G}^{\prime}}$, and we assume it is at most $L^{\prime}$.  Then, $\mathbf{G}^{\prime}$ can be modeled as follows:
\begin{equation}
\label{BS-RIS channel matrix}
{\mathbf{G}^{\prime}} = \sqrt{\frac{N_\text{s}N_\text{r}}{\rho^{\prime}}} \sum_{i=1}^{L^{\prime}} \alpha_{i} \mathbf{a}_{\mathrm{r}}(\gamma_{i}, \mu_{i}) \mathbf{a}_{\mathrm{s}}^{\mathrm{H}}(\theta_{i}, \phi_{i}) \text{,}
\end{equation}
where $\rho^{\prime}$ denotes the average path-loss between the BS and the RIS, $\alpha_{i}$ is the propagation gain associated with the $i^{\text{\textit{th}}}$ path, and $\mathbf{a}_{\mathrm{r}}(\gamma_{i}, \mu_{i})$ and $\mathbf{a}_{\mathrm{s}}(\theta_{i}, \phi_{i})$ are the array response vectors at the BS and RIS, respectively. We assume that the RIS deployed here is an $N_{\mathrm{r,h}} \times N_{\mathrm{r,w}}$ uniform planar array. Then, we have
\begin{equation}
\mathbf{a}_{\mathrm{s}}\left(\theta_{i}, \phi_{i}\right) = [e^{j(1-1)u_{\mathrm{s}}}, e^{j(2-1)u_{\mathrm{s}}}, \cdots, e^{j(N_{\mathrm{s}}-1)u_{\mathrm{s}}}]^{\mathrm{T}} \text{,} \quad \;\; 
\end{equation}
\begin{equation}
\label{3}
\begin{aligned}
\mathbf{a}_{\mathrm{r}}\left(\gamma_{i}, \mu_{i}\right) & =  \mathbf{a}_{\mathrm{r,h}}\left(\gamma_{i}, \mu_{i}\right) \otimes \mathbf{a}_{\mathrm{r,w}}\left(\gamma_{i}, \mu_{i}\right) \\
& = [e^{j(1-1)u_{\mathrm{r,h}}}, e^{j(2-1)u_{\mathrm{r,h}}}, \cdots, e^{j(N_{\mathrm{r,h}}-1)u_{\mathrm{r,h}}}]^{\mathrm{T}} \\
& \otimes [e^{j(1-1)u_{\mathrm{r,w}}}, e^{j(2-1)u_{\mathrm{r,w}}}, \cdots, e^{j(N_{\mathrm{r,w}}-1)u_{\mathrm{r,w}}}]^{\mathrm{T}}  \text{,}
\end{aligned}
\end{equation}
where $\otimes$ represents the Kronecker product, the directional  parameters: $u_{\mathrm{s}} = \frac{2\pi d}{\lambda}\sin(\theta_{i}) \cos (\phi_{i})$, $u_{\mathrm{r,h}} = \frac{2\pi d}{\lambda} \cos(\gamma_{i})$, and $u_{\mathrm{r,w}} = \frac{2\pi d}{\lambda}\sin(\gamma_{i}) \cos (\mu_{i})$, $d$ is the separation between antennas (reflecting elements) at the BS (RIS), and $\lambda$ is the wavelength of transmitted signal. Similarly, we assume that the number of paths between the RIS and MS is at most $L^{\prime\prime}$. Then, the RIS-MS channel matrix $\mathbf{G}^{\prime\prime}$ is modeled as follows:
\begin{equation}
\label{RIS-user channel matrix}
\mathbf{G}^{\prime\prime} = \sqrt{\frac{N_\mathrm{r}N_\mathrm{d}}{\rho^{\prime\prime}}} \sum_{i=1}^{L^{\prime\prime}} \beta_{i} \mathbf{a}_{\mathrm{d}}(\vartheta_{i}, \varphi_{i})  \mathbf{a}_{\mathrm{r}}^{\mathrm{H}}(\gamma^{\prime}_{i}, \mu^{\prime}_{i}) \text{,}
\end{equation}
where $\rho^{\prime\prime}$ denotes the average path-loss between the RIS and the user, $\beta_{i}$ is the propagation gain associated with the $i^{\text{\textit{th}}}$ path, and $\mathbf{a}_{\mathrm{d}}\left(\vartheta_{i}, \varphi_{i}\right)$ is the array response vector at the MS, which can be written as 
\begin{equation}
\label{5}
\mathbf{a}_{\mathrm{d}}\left(\vartheta_{i}, \varphi_{i}\right) = [e^{j(1-1)u_{\mathrm{d}}}, e^{j(2-1)u_{\mathrm{d}}}, \cdots, e^{j(N_{\mathrm{d}}-1)u_{\mathrm{d}}}]^{\mathrm{T}} \text{,}
\end{equation}
where $u_{\mathrm{d}} = \frac{2\pi d}{\lambda}\sin(\vartheta_{i}) \cos (\varphi_{i})$. Based on the BS-RIS and RIS-MS channel models established in (\ref{BS-RIS channel matrix}) and (\ref{RIS-user channel matrix}), the overall $N_{\mathrm{d}} \times N_{\mathrm{s}}$ channel matrix $\mathbf{H}$ can be expressed as 
\begin{equation}
	\label{channel matrix}
	{\mathbf{H}} = \mathbf{G}^{\prime\prime} \mathbf{\Phi} \mathbf{G}^{\prime} \text{,}
\end{equation}
where the diagonal matrix $\mathbf{\Phi}= \operatorname{diag}[e^{j\textcolor{black}{\boldsymbol{\varrho}}}]$ is the response at the RIS \footnote{
	Since the RIS is a passive device, each reflecting element is usually designed to maximize the signal reflection. Thus, we set the amplitude of reflection coefficient equal to one for simplicity in this letter.
}, and the $N_{\mathrm{r}}$ dimensional vector $\textcolor{black}{\boldsymbol{\varrho}=[\varrho_1, \cdots, \varrho_{N_{\mathrm{r}}}]^{\mathrm{T}}}$ represents the phase shifts of reflecting elements at the RIS. 

Then,  the received signals $\mathbf{Y} \in \mathbb{C}^{N_{\mathrm{s}} \times K}$ at the BS over $K$ time slots can be expressed as 
\begin{equation}
\label{received signal}
\begin{aligned}
\mathbf{Y} & = \mathbf{U}_{\mathrm{s}}^{\mathrm{H}} \left[ \mathbf{H}^{\mathrm{H}} \left( \mathbf{U}_{\mathrm{d}}\mathbf{X}\right)  + \mathbf{N}\right]\\
& = \mathbf{U}_{\mathrm{s}}^{\mathrm{H}} \mathbf{H}^{\mathrm{H}} \mathbf{U}_{\mathrm{d}}\mathbf{X} + \tilde{\mathbf{N}} \text{,}
\end{aligned}
\end{equation}
where $\mathbf{U}_{\mathrm{d}}$ and $\mathbf{U}_{\mathrm{s}}^{\mathrm{H}}$ are the transmit beamforming and receive combining matrices, respectively, $\mathbf{X}$ represents the pilot signal transmitted by the MS, $\tilde{\mathbf{N}}$ is the additive white Gaussian noise with the elements independently drawn from $\mathcal{C}\mathcal{N} (0, \sigma^{2})$. The $i^{\textit{th}}$ columns of $\mathbf{X}$ and $\tilde{\mathbf{N}}$ are corresponding to the $i^{\textit{th}}$ time slot, and we denote the transmit power as $p_{\mathrm{MS}} = \mathbb{E}\lbrace{\boldsymbol{x}^{\mathrm{H}}[i]}\boldsymbol{x}[i]\rbrace $.

\section{Sparse Structure of Cascade Channel}
\begin{figure}[htbp]
	\centerline{\includegraphics[width = 6.5 cm]{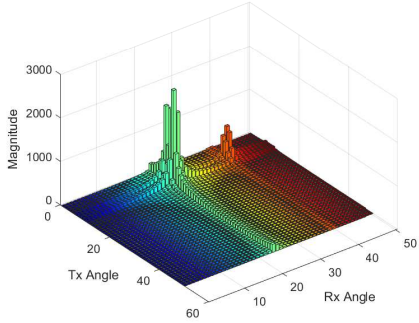}}
	\caption{Angular-domain channel for $N_{\mathrm{s}} = 50$, $N_{\mathrm{d}} = 50$, and $N_{\mathrm{r}}=40$. BS-RIS channel has $2$ paths and RIS-MS channel has $2$ paths.}
	\vspace{-10 pt}
	\label{fig 2}
\end{figure}

Before estimating the cascade channel $\mathbf{H}$, the first problem we are facing now is how to convert the estimation task into a noisy sparse signal recovery problem since the representation of $\mathbf{H}$ in (\ref{channel matrix}) is not visibly sparse.
To this end, pre-discretized grids can be utilized to establish the sparse representation {\cite{9103231}}.
However, this method may cause grid mismatch and estimation accuracy reduction. 
Another issue we should note is that even mildly ill-conditioned sensing matrices can lead to estimation failure in a compressed sensing problem {\cite{8698290, 6875146}}. 
In order to prevent these issues, we give the sparse representation by expressing the cascade channel in the angular domain based on suitable DFT bases.
Thus, the beamforming matrices $\mathbf{U}_{\mathrm{d}}$ and $\mathbf{U}_{\mathrm{s}}^{\mathrm{H}}$ are set as the $N_{\mathrm{d}} \times N_{\mathrm{d}}$ and $N_{\mathrm{s}} \times N_{\mathrm{s}}$ spatial unitary DFT matrices, respectively.
A given path with the directional parameters $u_{\mathrm{s}}$ and $u_{\mathrm{d}}$, which are defined under (\ref{3}) and (\ref{5}), has almost all of its energy along the particular vectors $[\mathbf{U}_{\mathrm{s}}]_{:,m}$ and $[\mathbf{U}_{\mathrm{d}}]_{:,n}$, and very little along all the others, if $m$ and $n$ satisfy {\cite{8611231}}:
\begin{equation}
\left|u_{\mathrm{s}} - \frac{2\pi(m-1)}{N_{\mathrm{s}}} \right| < \frac{2\pi}{N_{\mathrm{s}}} \text{,}
\end{equation}
\begin{equation}
\left|u_{\mathrm{d}} - \frac{2\pi(n-1)}{N_{\mathrm{d}}} \right| < \frac{2\pi}{N_{\mathrm{d}}} \text{.}
\end{equation}
In order to illustrate visually, \textcolor{black}{Fig. \ref{fig 2}} plots a specific realization for the channel magnitude in the angular domain. As seen from it, the true channel is indeed sparse in the angular domain, i.e., it exhibits a few dominant coefficients.
Consequently, the RIS-assisted mmWave channel is inherently sparse in the angular domain if expressed in suitable DFT bases.
 
Utilizing the DFT beamforming matrices $\mathbf{U}_{\mathrm{d}}$ and $\mathbf{U}_{\mathrm{s}}^{\mathrm{H}}$ and vectorizing the received signals $\mathbf{Y}$ at the BS yields
\begin{equation}
\label{13}
\begin{aligned}
\boldsymbol{y}& = \operatorname{vec}\left\lbrace\mathbf{U}_{\mathrm{s}}^{\mathrm{H}} \mathbf{H}^{\mathrm{H}} \mathbf{U}_{\mathrm{d}} \mathbf{X} + \tilde{\mathbf{N}}\right\rbrace = \operatorname{vec} (\tilde{\mathbf{H}}^{\mathrm{H}} \mathbf{X})  + \operatorname{vec} (\tilde{\mathbf{N}}) \\
& \overset{(a)}{=}  {\left(\mathbf{X}^{\mathrm{T}} \otimes \mathbf{I}_{N_{\mathrm{s}}} \right)} {\operatorname{vec} (\tilde{\mathbf{H}}^{\mathrm{H}})} + \operatorname{vec} (\tilde{\mathbf{N}}) = {\mathbf{\Upsilon}}{\boldsymbol{\upsilon}} + \boldsymbol{n} \text{,}
\end{aligned}
\end{equation}
where $\tilde{\mathbf{H}}^{\mathrm{H}} = \mathbf{U}_{\mathrm{s}}^{\mathrm{H}} \mathbf{H}^{\mathrm{H}} \mathbf{U}_{\mathrm{d}}$ is the cascade channel represented in the angular domain, $\boldsymbol{\upsilon} = \operatorname{vec} (\tilde{\mathbf{H}}^{\mathrm{H}})$ is the sparse signal that we need to recover, $\mathbf{\Upsilon} = \mathbf{X}^{\mathrm{T}} \otimes \mathbf{I}_{N_{\mathrm{s}}}$ denotes the measurement matrix, $\boldsymbol{n} = \operatorname{vec} (\tilde{\mathbf{N}})$ is the additive Gaussian noise, and the equality (a) follows from the relation of the vectorization of the matrix product to the Kronecker product {\cite{zhang2017matrix}}. We assume that $\boldsymbol{\upsilon}$ is sparse with at most $L \propto L^{\prime} \times L^{\prime\prime}$ non-zero entries in unknown locations. The sparse-level $L$ is actually a prior information and is related to the number of paths. Once $\boldsymbol{\upsilon}$ is recovered, an estimate of $\mathbf{H}$ is readily obtained as follows:
\begin{equation}
\hat{\mathbf{H}}
= \mathbf{U}_{\mathrm{d}} \hat{\operatorname{\tilde{\mathbf{H}}}} \mathbf{U}_{\mathrm{s}}^{\mathrm{H}} \text{,}
\end{equation}
where $\hat{\operatorname{\tilde{\mathbf{H}}}}^{\mathrm{H}} = \operatorname{unvec}\left(\hat{\boldsymbol{\upsilon}}\right)$ and $\hat{\boldsymbol{\upsilon}}$ is an estimate of ${\boldsymbol{\upsilon}}$. Moreover, the estimate of $\operatorname{vec} ({\mathbf{H}}) = (\mathbf{G}^{\prime \mathrm{T}} \otimes \mathbf{G}^{\prime\prime}) \operatorname{vec}(\mathbf{\Phi})$ {\cite{zhang2017matrix}} is enough to
configure the phase shifts at RIS because the beamforming problem can be converted to an optimization problem which maximizes $ \|\mathbf{H}\|^{2}_{\mathrm{F}} = \|\operatorname{vec}\left( \mathbf{H}\right)\|^{2}_{2}$ with respect to $\operatorname{vec}(\mathbf{\Phi})$.

\section{Asymptotic Achievability of The Cramér-Rao Lower Bound via Joint Typicality Estimator}

Many classical compressed sensing algorithms such as basis pursuit (BP) {\cite{1614066}} and orthogonal matching pursuit (OMP) {\cite{4385788}} can be utilized to recover the sparse signal $\boldsymbol{\upsilon}$. However, these algorithms always choose the locally optimal approximation to the actual sparse signal {\cite{1614066, 4385788, 5550495, 5419092, 4839056}}. Thus, in this section, we utilize the Shannon theory and the notion of joint typicality {\cite{4694104}} to asymptotically achieve the CRLB of the channel estimation for RIS-assisted mmWave systems where the estimator has no knowledge of the actual locations of the non-zero entries in $\boldsymbol{\upsilon}$. To prove the asymptotic achievability of the CRLB, we first state the following lemma.

\begin{lemma}
\label{fullrank}
Let the set $\mathcal{J} \subset \left\lbrace 1, \cdots, N_{\mathrm{d}}N_{\mathrm{s}} \right\rbrace $ such that $|\mathcal{J}| = L$ and $\mathbf{\Upsilon}_{\mathcal{J}}$ be the sub-matrix of the measurement matrix $\mathbf{\Upsilon}$ with the columns corresponding to the index set $ \mathcal{J}$. Then, we have $\operatorname{rank}(\mathbf{\Upsilon}_{\mathcal{J}})= L$ with probability $1$.
\end{lemma}

\begin{IEEEproof}
First, we consider the rank of $\mathbf{X}^{\mathrm{T}}$. The $(m,n)^{\textit{th}}$ entry of it represents the pilot symbol transmitted by the $n^{\textit{th}}$ antenna at the $m^{\textit{th}}$ time slot. Thus, all of the entries in it are independent and designable. For simplicity, we set them as independent and identically distributed (i.i.d.) and distributed according to $\mathcal{CN}(0,1)$. Let $\boldsymbol{x}_{i}$ and $\boldsymbol{x}_{j}$ be two columns of $\mathbf{X}^{\mathrm{T}}$. Utilizing the law of large numbers yields
\begin{equation}
\boldsymbol{x}_{i}^{\mathrm{H}} \boldsymbol{x}_{j} = \sum_{k} x_{k,i}^{*}x_{k,j} \rightarrow 0  \text{,} \;\; i \neq j \text{,}
\end{equation}
as $K$ goes to infinity. Thus, the columns of $\mathbf{X}^{\mathrm{T}}$ are mutually orthogonal with probability $1$, i.e., $\mathbf{X}^{\mathrm{T}}$ is a full column rank matrix when $K>N_{\mathrm{d}}$. 
Then, due to $\mathbf{I}_{N_{\mathrm{s}}}$ is a unit matrix, it has a full column rank.
By utilizing the rank property of the Kronecker product:  $\operatorname{rank}(\boldsymbol{\Upsilon}) = \operatorname{rank}(\mathbf{X}^{\mathrm{T}})\operatorname{rank}(\mathbf{I}_{N_{\mathrm{s}}})$, we prove the statement of this lemma.	
\end{IEEEproof}

\vspace{2 pt}
Then, to establish the joint typicality-based channel estimator, we need to define the notion of joint typicality. We adopt the definition from {\cite{4694104}} which is given as follows: 
\vspace{- 2 pt}
\begin{definition}{($\delta$-Jointly Typicality)}
	\label{Joint Typicality}
	\item The received signal ${\boldsymbol{y}}$ collected over $K$ time slots, and the set of indices $\mathcal{J} \subset \left\lbrace 1, 2, \cdots, N_{\mathrm{d}}N_{\mathrm{s}} \right\rbrace $ with $|\mathcal{J}| = L$ are $\delta$-jointly typical, if $\operatorname{rank}(\mathbf{\Upsilon}_{\mathcal{J}})= L$ and 
	\begin{equation}
	\left| \frac{1}{K N_{\mathrm{s}}}\| \mathbf{\Pi}_{\mathbf{\Upsilon}_{\mathcal{J}}}^{\perp} {\boldsymbol{y}} \|^{2}-\frac{K N_{\mathrm{s}}-L}{K N_{\mathrm{s}}} \sigma^{2} \right| < \delta \text{,}
	\end{equation}
	where $\mathbf{\Upsilon}_{\mathcal{J}}$ is the sub-matrix of the measurement matrix $\mathbf{\Upsilon}$ with the columns corresponding to the index set $ \mathcal{J}$, and $\mathbf{\Pi}_{\mathbf{\Upsilon}_{\mathcal{J}}}^{\perp} = \mathbf{I}-\mathbf{\Upsilon}_{\mathcal{J}}(\mathbf{\Upsilon}_{\mathcal{J}}^{\mathrm{H}} \mathbf{\Upsilon}_{\mathcal{J}})^{-1}\mathbf{\Upsilon}_{\mathcal{J}}^{\mathrm{H}}$ is the orthogonal projection matrix.
\end{definition}

Next, we establish the following proposition to show that the proposed estimator can be applied to the considered problem.

\begin{proposition}
	The joint typicality-based estimator can be utilized to estimate the cascade channel in an RIS-assisted mmWave system, \textit{i.e.}, solve the noisy sparse signal recovery problem in Eq. (\ref{13}). The detailed channel estimation steps are illustrated in Algorithm 1.
\end{proposition}

\vspace{-2 pt}
\begin{IEEEproof} 
	The measurement matrix $\boldsymbol{\Upsilon}$ in (\ref{13}) is proved to be full column rank in Lemma \ref{fullrank}, which ensures that the sub-spaces spanned by different $L$ column vectors chosen from the measurement matrix $\boldsymbol{\Upsilon}$ are different. Based on Definition \ref{Joint Typicality}, if $L$ column vectors are chosen correctly, there exists only additive white Gaussian noise in the orthogonal complement. Thus, the joint typicality-based estimator can be utilized to solve the noisy sparse signal recovery problem in (\ref{13}). 
\end{IEEEproof}
\vspace{2 pt}

\begin{algorithm}[!h]
	\caption{Joint Typicality-Based Channel Estimator}
	\begin{algorithmic}[1]
		\STATE \textbf{Input:} The numbers of antennas $N_{\mathrm{s}}$ at the BS and $N_{\mathrm{d}}$ at the MS, the pilot signal $\mathbf{X}$, the received signal vector ${\boldsymbol{y}}$, and the maximal sparse-level $L$.
		
		\WHILE{index set $\mathcal{J}_{i-1}$ is not $\delta$-jointly typical with ${\boldsymbol{y}}$}
		\STATE $i^{\text{\textit{th}}}$ \textit{iteration of all the possible $\binom{N_{\mathrm{d}}N_{\mathrm{s}}}{L}$ $L$-dimensional sub-spaces }:
		\STATE Determine whether the following inequality is satisfied.
		\begin{equation*}
		\left| \frac{1}{K N_{\mathrm{s}}}\| \mathbf{\Pi}_{\mathbf{\Upsilon}_{\mathcal{J}_{i}}}^{\perp} {\boldsymbol{y}} \|^{2}-\frac{K N_{\mathrm{s}}-L}{K N_{\mathrm{s}}} \sigma^{2} \right| < \delta
		\end{equation*}
		\STATE If it is satisfied, compute the estimate $\hat{\boldsymbol{\upsilon}}$ by projecting the received signal ${\boldsymbol{y}}$ onto the sub-space spanned by $\mathbf{\Upsilon}_{\mathcal{J}_{i}}$.
		\begin{equation*}
		\hat{\boldsymbol{\upsilon}} = (\mathbf{\Upsilon}_{\mathcal{J}_{i}}^{\mathrm{H}} \mathbf{\Upsilon}_{\mathcal{J}_{i}})^{-1}\mathbf{\Upsilon}_{\mathcal{J}_{i}}^{\mathrm{H}}{\boldsymbol{y}}
		\end{equation*}
		\ENDWHILE
		\STATE If there exists no set that is $\delta$-jointly typical to $\boldsymbol{y}$, it outputs the zero vector.
		\STATE \textbf{Output:} The channel estimate $\hat{\mathbf{H}}^{\mathrm{H}} = \mathbf{U}_{\mathrm{s}} \operatorname{unvec}(\hat{\boldsymbol{\upsilon}})\mathbf{U}_{\mathrm{d}}^{\mathrm{H}}$.
	\end{algorithmic}
\end{algorithm}

In order to further prove that we can asymptotically achieve the CRLB on the estimation error where the estimator has no knowledge of the locations of the non-zero entries in $\boldsymbol{\upsilon}$, we state the following lemmas.
\vspace{-2 pt}
\begin{lemma}
	\label{CRLBlemma}
	For any unbiased estimate $\hat{\boldsymbol{\upsilon}}$ of $\boldsymbol{\upsilon}$, the Cramér-Rao lower bound on the MSE is given as
	\begin{equation}
	\label{CRLB-Lambda}
	\mathbb{E\left\lbrace \| \hat{\boldsymbol{\upsilon}} - \boldsymbol{\upsilon}\|^{\mathrm{2}} \right\rbrace } \geq \sigma^{2} \operatorname{Tr}\left[(\mathbf{\Upsilon}_{\mathcal{I}}^{\mathrm{H}} \mathbf{\Upsilon}_{\mathcal{I}})^{-1}\right] \text{.}
	\end{equation}
\end{lemma}

\begin{IEEEproof}
	The likelihood function of the random vector ${\boldsymbol{y}}$ conditioned on $\boldsymbol{\upsilon}$ is
	\begin{equation}
	p({\boldsymbol{y}}; \boldsymbol{\upsilon}) = \frac{\operatorname{exp}\left( -\frac{1}{2\sigma^2} \|{\boldsymbol{y}} - \mathbf{\Upsilon}_{\mathcal{I}} \boldsymbol{\upsilon}_{\mathcal{I}}\|^{2}\right)}{(2\pi)^{K N_{\mathrm{s}}/2}\sigma^{K N_{\mathrm{s}}}} \text{,}
	\end{equation}
	where $\boldsymbol{\upsilon}_{\mathcal{I}}$ is the subvector of $\boldsymbol{\upsilon}$  with elements corresponding to the index set $\mathcal{I}$. Then, by using (6) in {\cite{1519691}}, the CRLB can be written as (\ref{CRLB-Lambda}).
\end{IEEEproof}

\vspace{-2 pt}
\begin{lemma}{\rm(Lemma 2.3 of {\cite{5361481}})}
	\label{lemma 2.3 of shannon limit}
	\\ Let $\mathcal{I} = \operatorname{supp}(\boldsymbol{\upsilon})$ and $\operatorname{rank}(\mathcal{\mathbf{\Upsilon}_{\mathcal{I}}}) = L$. Then, for $\delta > 0$, it holds that
	\begin{equation}
	\begin{aligned}
	& \mathbb{P}\left( \left| \frac{1}{K N_{\mathrm{s}}}\| \mathbf{\Pi}_{\mathbf{\Upsilon}_{\mathcal{J}}}^{\perp} {\boldsymbol{y}} \|^{2}-\frac{K N_{\mathrm{s}}-L}{K N_{\mathrm{s}}} \sigma^{2} \right| > \delta \right) \\ & \quad \leq 2 \operatorname{exp}\left( {-\frac{\delta^{2}}{4 \sigma^{4}} \frac{K^{2} N_{\mathrm{s}}^{2}}{K N_{\mathrm{s}}-L+\frac{2\delta}{\sigma^{2}}K N_{\mathrm{s}} }} \right) \text{.}
	\end{aligned}
	\end{equation}	
	Let $\mathcal{J}$ be an index set such that $|\mathcal{J}|= L$, $|\mathcal{I} \cap \mathcal{J}| < L$, and $\operatorname{rank}(\mathbf{\Upsilon}_{\mathcal{J}}) = L$. Then, for $\delta > 0$, it holds that
	\begin{equation}
	\begin{aligned}
	&\mathbb{P}\left( \left| \frac{1}{K N_{\mathrm{s}}}\| \mathbf{\Pi}_{\mathbf{\Upsilon}_{\mathcal{J}}}^{\perp} {\boldsymbol{y}} \|^{2}-\frac{K N_{\mathrm{s}}-L}{K N_{\mathrm{s}}} \sigma^{2} \right| < \delta \right) \\ 
	& \quad \leq \operatorname{exp}\left( \frac{{L - K N_{\mathrm{s}}}}{4} \left( \frac{\sum_{k\in\mathcal{I} \backslash \mathcal{J}}{|{\upsilon}_{k}|^{2}}-\delta^{\prime}}{\sum_{k\in\mathcal{I} \backslash \mathcal{J}}{|{\upsilon}_{k}|^{2}}+\sigma^{2}} \right)^{2} \right) \text{,}
	\end{aligned} 
	\end{equation}
	where ${\upsilon}_{k}$ is the $k^{\textit{th}}$ entry in $\boldsymbol{\upsilon}$ and
	\begin{equation}
	\delta^{\prime} = \delta \frac{K N_{\mathrm{s}}}{K N_{\mathrm{s}}- L} \text{.}
	\end{equation}	
\end{lemma}
\begin{IEEEproof}
	Please refer to {\cite{5361481}} for the proof.
\end{IEEEproof}

\vspace{5 pt}
Finally, based on the above lemmas, we establish the asymptotic achievability of the CRLB in the following theorem.
\vspace{-2 pt}
\begin{theorem}
	\label{theorem 2}
	By utilizing the joint typicality-based channel estimator given in Algorithm 1, the MSE of cascade channel estimation in an RIS-assisted mmWave system asymptotically achieves the CRLB as the product of the number of receiver antennas and the number of time slots tends to infinity. This bound can be asymptotically achieved whether the estimator knows the location of the non-zero entries.
\end{theorem}

\begin{IEEEproof}
	The MSE of the joint typicality estimator (averaged over all possible measurement matrices) can be upper-bounded as follows:
	\begin{equation*}
	\begin{aligned}
	\varepsilon_{\delta}(K N_{\mathrm{s}}) = & \mathbb{E}\left\lbrace \|\hat{\boldsymbol{\upsilon}}-\boldsymbol{\upsilon} \|^{2} \right\rbrace \\ \leq & \int_{\boldsymbol{\Upsilon}} \|\boldsymbol{\upsilon}\|^{2} \mathbb{P}(\mathrm{E}_{0}) dP(\boldsymbol{\Upsilon}) \\  + & \int_{\boldsymbol{\Upsilon}} \mathbb{E}_{{\boldsymbol{n}}|\boldsymbol{\Upsilon}} \left\lbrace \| (\mathbf{\Upsilon}_{\mathcal{I}}^{\mathrm{H}} \mathbf{\Upsilon}_{\mathcal{I}})^{-1}\mathbf{\Upsilon}_{\mathcal{I}}^{\mathrm{H}} {\boldsymbol{y}} - \boldsymbol{\upsilon}\|^{2} \right\rbrace \\ & \quad \; \times \mathbb{P}(\mathcal{I} \sim {\boldsymbol{y}}) dP(\boldsymbol{\Upsilon}) 
	\end{aligned}
	\end{equation*}
	\begin{equation}
	\label{MSEjoint}
	\begin{aligned}
	+ & \int_{\boldsymbol{\Upsilon}} \sum_{\mathcal{J} \neq \mathcal{I}} \mathbb{E}_{{\boldsymbol{n}}|\boldsymbol{\Upsilon}} \left\lbrace \| (\mathbf{\Upsilon}_{\mathcal{J}}^{\mathrm{H}} \mathbf{\Upsilon}_{\mathcal{J}})^{-1}\mathbf{\Upsilon}_{\mathcal{J}}^{\mathrm{H}} {\boldsymbol{y}} - \boldsymbol{\upsilon}\|^{2} \right\rbrace \\ & \quad \; \times \mathbb{P}(\mathcal{J} \sim {\boldsymbol{y}}) dP(\boldsymbol{\Upsilon}) \text{,}
	\end{aligned}
	\end{equation}
	where $\mathbb{P}(\cdot)$ represents the event probability defined over the noise density, the event $\mathrm{E}_{0}$ represents the estimator does not find any set $\delta$-jointly typical to ${\boldsymbol{y}}$, $dP(\boldsymbol{\Upsilon})$ represents the probability measure of the matrix $\boldsymbol{\Upsilon}$, and the inequality follows from the Boole’s inequality. The second term is corresponding to $\mathcal{I}$ and is the MSE of a genie-aided estimation where the estimator knows $\operatorname{supp}(\boldsymbol{\upsilon})$. We rewrite it as follows:
	\begin{equation}
	\begin{aligned}
	& \int_{\boldsymbol{\Upsilon}} \mathbb{E}_{{\boldsymbol{n}}|\boldsymbol{\Upsilon}} \left\lbrace \| (\mathbf{\Upsilon}_{\mathcal{I}}^{\mathrm{H}} \mathbf{\Upsilon}_{\mathcal{I}})^{-1}\mathbf{\Upsilon}_{\mathcal{I}}^{\mathrm{H}}{\boldsymbol{y}} - \boldsymbol{\upsilon}\|^{2} \right\rbrace \mathbb{P}(\mathcal{I} \sim {\boldsymbol{y}}) dP(\boldsymbol{\Upsilon}) \\ 
	& = \mathbb{E}_{{\boldsymbol{n}}, \boldsymbol{\Upsilon}} \left\lbrace \| (\mathbf{\Upsilon}_{\mathcal{I}}^{\mathrm{H}} \mathbf{\Upsilon}_{\mathcal{I}})^{-1}\mathbf{\Upsilon}_{\mathcal{I}}^{\mathrm{H}} {\boldsymbol{n}}\|^{2} \right\rbrace
	= \mathbb{E}_{\boldsymbol{\Upsilon}} \left\lbrace \sigma^{2} \operatorname{Tr}(\mathbf{\Upsilon}_{\mathcal{I}}^{\mathrm{H}} \mathbf{\Upsilon}_{\mathcal{I}})^{-1}\right\rbrace \text{.}
	\end{aligned}
	\end{equation}
	By using Lemma \ref{CRLBlemma}, we obtain that the second term in (\ref{MSEjoint}) is the CRLB of the genie-aided cascade channel estimation.
	
	Next, we show that the first and third term in (\ref{MSEjoint}) converge to zero when $K N_{\mathrm{s}} \rightarrow \infty$. By using Lemma \ref{lemma 2.3 of shannon limit}, the first term can be upper-bounded as 
	\begin{equation}
	\begin{aligned}
	& \int_{\boldsymbol{\Upsilon}} \|\boldsymbol{\upsilon}\|^{2} \mathbb{P}(\mathrm{E}_{0}) dP(\boldsymbol{\Upsilon}) \\ & \quad \leq 2 \|\boldsymbol{\upsilon}\|^{2} \operatorname{exp}\left( {-\frac{\delta^{2}}{4 \sigma^{4}} \frac{K^{2}N_{\mathrm{s}}^{2}}{K N_{\mathrm{s}}-L+\frac{2\delta}{\sigma^{2}}KN_{\mathrm{s}}}} \right) \text{.}
	\end{aligned}
	\end{equation}
	This term approaches to zero as $K N_{\mathrm{s}} \rightarrow \infty$, since $\|\boldsymbol{\upsilon}\|^{2}$ grows polynomially in $ N_{\mathrm{s}} $ and the exponential term tends to negative infinity as $K N_{\mathrm{s}} \rightarrow \infty$. By using Lemma \ref{lemma 2.3 of shannon limit}, the third term can be upper-bounded as 
	\begin{equation}
	\label{term 3}
	\begin{aligned}
	& \int_{\boldsymbol{\Upsilon}} \sum_{\mathcal{J} \neq \mathcal{I}} \mathbb{E}_{{\boldsymbol{n}}|\boldsymbol{\Upsilon}} \left\lbrace \| (\mathbf{\Upsilon}_{\mathcal{J}}^{\mathrm{H}} \mathbf{\Upsilon}_{\mathcal{J}})^{-1}\mathbf{\Upsilon}_{\mathcal{J}}^{\mathrm{H}} {\boldsymbol{y}} - \boldsymbol{\upsilon}\|^{2} \right\rbrace \\ & \quad \quad \quad \; \times \mathbb{P}(\mathcal{J} \sim {\boldsymbol{y}}) dP(\boldsymbol{\Upsilon}) \\
	&\quad \leq (L \sigma^{2}+\|\boldsymbol{\upsilon}\|^{2}) \int_{\boldsymbol{\Upsilon}} \sum_{\mathcal{J} \neq \mathcal{I}} \mathbb{E}_{{\boldsymbol{n}}|\boldsymbol{\Upsilon}} \mathbb{P}(\mathcal{J} \sim {\boldsymbol{y}}) dP(\boldsymbol{\Upsilon}) \\ 
	& \quad \leq (L \sigma^{2}+\|\boldsymbol{\upsilon}\|^{2}) \; \times \\ & \quad \; \quad \sum_{\mathcal{J} \neq \mathcal{I}}\operatorname{exp}\left( \frac{{L-K N_{\mathrm{s}}}}{4} \left( \frac{\sum_{k\in\mathcal{I} \backslash \mathcal{J}}{|{\upsilon}_{k}|^{2}}-\delta^{\prime}}{\sum_{k\in\mathcal{I} \backslash \mathcal{J}}{|{\upsilon}_{k}|^{2}}+\sigma^{2}} \right)^{2} \right) \text{.}
	\end{aligned}
	\end{equation}
	This term tends to zero as $K N_{\mathrm{s}} \rightarrow \infty$, since $(L \sigma^{2}+\|\boldsymbol{\upsilon}\|^{2})$ grows polynomially in $ N_{\mathrm{s}} $ and $(L - K N_{\mathrm{s}})$ tends to negative infinity as $K N_{\mathrm{s}} \rightarrow \infty$.
\end{IEEEproof}

\section{Numerical Results}

\begin{figure}[htbp]
	\centerline{\includegraphics[width = 9 cm]{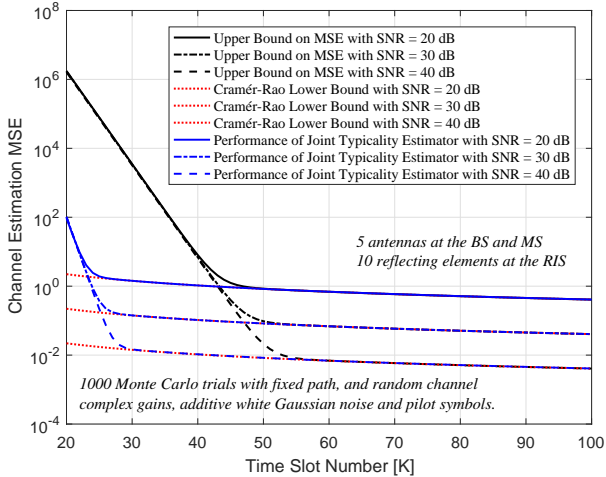}}
	\caption{The performance of joint typicality-based channel estimator versus the time slot number with different SNRs.}
	\label{fig 3}
\end{figure}

In this section, we numerically illustrate the result given in Theorem \ref{theorem 2}. To verify whether the CRLB of cascade channel estimation for RIS-assisted mmWave communication systems can be asymptotically achieved when the product of time slot number and receiver antenna number $K N_{\mathrm{s}}$ tends to infinity, {Fig. \ref{fig 3}} simultaneously plots the curves of the CRLB, the MSE upper bound, and the performance of joint typicality estimator versus the time slot number $K$ with different signal-to-noise ratios (SNRs) selected from the set of $\left\lbrace20\text{ dB}, 30\text{ dB}, 40\text{ dB} \right\rbrace$. 
In this figure, the numbers of antennas at the BS and the MS are both set as $5$, and the number of reflecting elements at the RIS is set as $10$. The path numbers in the BS-RIS channel and the RIS-MS channel are both set as $1$. In addition, the numerical results in {Fig. \ref{fig 3}} are obtained through $1,000$ Monte Carlo trials. It is observed that the CRLB can be achieved as the time slot number tends to infinity, which confirms the result in Theorem \ref{theorem 2}. When we fix the time slot number $K$ and change receiver antenna number $N_{\mathrm{s}}$, the curves are similar to {Fig. \ref{fig 3}}, and we omit it due to the space limitation. It is encouraging not only because the CRLB of cascade channel estimation for RIS-assisted mmWave systems can be asymptotically achieved but also because we can decrease the number of time slots consumed in channel estimation through increasing the number of receiver antennas.

\section{Conclusion}
In this letter, we consider the estimation of the cascade channel in an RIS-assisted mmWave communication system. By utilizing the joint typicality-based channel estimator, the MSE of estimation can asymptotically achieve the CRLB as the product of the number of receiver antennas and the number of time slots tends to infinity, and this bound can be asymptotically achieved whether the estimator knows the locations of the non-zero entries. To the best of our knowledge, it is the first research which establishes the asymptotic achievability of the CRLB of the cascade channel estimation for the RIS-assisted mmWave systems. Our result also reveals that the training overhead can be reduced through deploying more receiver antennas. However, there is an important issue that our established scheme is complex and costs a lot of overhead, thus finding a lower-complexity estimator that can simultaneously
achieve the CRLB for RIS-assisted mmWave systems is an important work in future studies.

\bibliographystyle{IEEEtran} 
\bibliography{reference}

\end{document}